\documentclass{llncs}

\title{A Delta Debugger for ILP Query Execution}
\author{Remko Tron\c{c}on\thanks{Supported by the Institute for the Promotion of
                    Innovation by Science and Technology in Flanders (I.W.T.)}
		 \and Gerda Janssens}
        
\institute{Katholieke Universiteit Leuven, Dept. of Computer Science,\\
    Celestijnenlaan 200A, B-3001 Leuven, Belgium\\
    \email{\{remko,gerda\}@cs.kuleuven.be}
}

\usepackage{alltt}
\usepackage{graphicx}

\newcommand{\href}[2]{#2}
\newcommand{\p}[1]{\texttt{#1}}
\newcommand{\Tilde}{\textsc{Tilde}}

\begin{document}

\maketitle

\begin{abstract}
Because query execution is the most crucial part of Inductive Logic
Programming (ILP) algorithms, a lot of effort is invested in developing faster
execution mechanisms. These execution mechanisms typically have a low-level
implementation, making them hard to debug. Moreover, other factors such as
the complexity of the problems handled by ILP algorithms and size
of the code base of ILP data mining systems make debugging
at this level a very difficult job. In this work, we present the
trace-based debugging
approach currently used in the development of new execution mechanisms
in hipP, the engine underlying the ACE Data Mining system. This debugger
uses the delta debugging algorithm to automatically reduce the total time 
needed to expose bugs in ILP execution, thus making manual debugging
step much lighter.
\end{abstract}

\section{Introduction}
Data mining \cite{Fayyad1996} is the process of finding patterns that
describe a large set of data best. Inductive Logic Programming (ILP) 
\cite{Muggleton1994} is a
multi-relational data mining approach, which uses the Logic Programming
paradigm as its
basis. ILP uses a generate-and-test approach, where in each iteration a
large set of hypotheses (or `queries') has to be evaluated on the data (also
called `examples'). Based on the results of this evaluation, the ILP 
process selects the ``best'' hypotheses and refines them further. 
Due to the size of the data of the problems handled by ILP, 
the underlying query evaluation engine (e.g. a Prolog system) is a
crucial part
of a real life ILP system. Hence, a lot of effort is invested in optimizing
the engine to yield faster evaluation time through the use of new execution
mechanisms, different internal data representations, etc.

The development of new execution mechanisms for ILP happens mainly in the
engine of the ILP system. These optimized execution strategies typically
require a low level implementation to yield significant benefits.
For example, the query pack~\cite{Blockeel2002} and adpack~\cite{Troncon2003b} 
execution mechanisms
require the introduction of new dedicated WAM instructions, together with
a set of new data structures which these instructions use and manipulate. 
Because of their low-level nature, finding bugs 
in the implementation of these execution mechanisms is very hard.
While tracing bugs in these low-level implementations 
might still be feasible for small test programs, many  bugs 
only appear during the execution of the ILP algorithm on real life data sets.
Several factors make debugging in this situation difficult:
\begin{itemize}
\item \emph{The size of the ILP system itself.} Real life ILP systems
      group the implementation of many algorithms into
      one big system. These systems therefore often have a very large code base. 
      For example, the ACE system~\cite{ace} consists of
      over 150000 lines of code. In the case of the ACE system, the code
	  base is very heterogeneous, where parts of code are written in different
	  languages and others are generated automatically using preprocessors etc.
	  This makes it in practice very hard to use standard tracing to detect 
      bugs.
\item \emph{The complexity/size of the ILP problem.} With large datasets, 
      it can take a very long
      time (hours, even days) before a specific bug occurs. When debugging,
      one typically performs multiple 
      runs with small modifications to pin-point the exact problem, and
      so long execution times make this approach infeasible.
\item \emph{The high complexity of the hypothesis generation phase.} While the
      evaluation of hypotheses is often the bottleneck, some algorithms 
      (such as rule learners) have a very expensive hypothesis generation 
      phase. This phase is independent from the execution of the queries
	  itself, and as such has no influence on the exposure of the bug.
	  For algorithms with a very complex hypothesis generation, it can take a 
	  very
      long time for the bug in the execution mechanism to expose itself,
      even when the time spent on executing these queries is small.
\item \emph{Non-determinacy of ILP algorithms.} If an ILP algorithm makes 
      random decisions (typically in the hypothesis generation phase), the 
	  exact point in time where the bug
      occurs changes from run to run. It is even possible that the bug
      does not occur at all in certain runs.
\end{itemize}

In \cite{Troncon2005}, we proposed a trace-based approach for analyzing
and debugging ILP data mining execution. This approach allowed
easy and fast debugging of the underlying query execution engines, 
independent of the ILP algorithm causing the bug to appear.
In this work, we present an extension to this debugging approach,
automating a large part of the debugging process. By applying the
\emph{delta debugging algorithm}~\cite{Zeller2002} on ILP execution traces,
we automatically generate minimal traces exposing a bug, thus greatly reducing
the time and effort needed to track the bug down. This approach is
currently used in the development of new execution mechanisms 
in hipP~\cite{hipp}, the engine underlying the ACE Data Mining
system~\cite{ace}. \\

The organization of this paper is as follows: In Section~\ref{sec:background},
we give a brief introduction to Inductive Logic Programming.
Section~\ref{sec:traces} discusses the collection of
the run-time information necessary for our trace-based debugging approach.
Section~\ref{sec:ddebug} then discusses applying the delta debugging algorithm
on these traces to allow fast and easy debugging. We briefly discuss the 
implementation of our delta debugger in Section~\ref{sec:implementation}.
Finally, we conclude in Section~\ref{sec:conclusion}.


\section{Background: Inductive Logic Programming} \label{sec:background}
The goal of Inductive Logic Programming is to find
a theory that best explains a large set of data (or examples).
More specifically, in the ILP setting at hand, 
each example is a logic program, and the theory is represented as a set 
of logical queries. Additionally, background knowledge about the problem
domain is encoded as logical predicates.

\begin{figure}[tbp]
\begin{tabbing}
\hspace{3mm} \= \hspace{3mm} \kill
\emph{\% QH: Queue of hypotheses} \\
$QH$ := \texttt{Initialize} \\
\textbf{repeat} \\
\> \texttt{Remove} $H$ from $QH$ \\
\> \texttt{Choose} refinements $r_1,\ldots,r_k$  to apply on $H$ \\
\> Apply refinements $r_1,\ldots,r_k$ on $H$ to get $H_1,\ldots,H_k$ \\
\> Add $H_1,\ldots,H_k$ to $QH$ \\
\> Evaluate $QH$ \\
\> \texttt{Prune} $QH$ \\
\textbf{until} \texttt{Stop-criterion}(QH)
\end{tabbing}
\caption{Generic ILP Algorithm} \label{fig:ilp}
\end{figure}
ILP algorithms typically follow the same generate-and-test pattern: a set of
queries is generated, of which all queries are tested on (a subset of) the
examples in the data set. The query (or queries) which cover the data the
best are then selected and extended, after which the process restarts with
the extended queries. Hence, the query space is exhaustively searched,
starting from the most general query and refining it further and further
until the query (or queries) cover the data well enough. The generic ILP
algorithm (as described in \cite{Muggleton1994}) can be seen in
Figure~\ref{fig:ilp}. In this algorithm, the \texttt{Initialize},
\texttt{Remove}, \texttt{Choose}, \texttt{Prune} and
\texttt{Stop-criterion} procedures are to be filled in by the ILP algorithm,
creating a special instance of this generic algorithm. Hence, these are the
functions that characterize an ILP algorithm.
In general, the most crucial step with respect to execution time is the
\textit{Evaluate} step: the (often large) set of 
queries $H_1,\ldots,H_n$ has to be run for
each example. It is not uncommon to have a set of 3000
queries which are executed up to 1000 times. Therefore, fast query execution
mechanisms are needed. Examples of these optimized techniques are 
query packs~\cite{Blockeel2002}, adpacks~\cite{Troncon2003b}, 
(lazy) control flow compilation~\cite{Troncon2006}, \ldots 
All of these techniques require a low-level implementation in the engine
that the ILP algorithm uses. Due to the low-level nature of these optimized
execution mechanisms, bugs in them are very hard to trace. 

Concrete examples of ILP algorithms are \textsc{Tilde} \cite{Blockeel1998}, 
a decision tree learner, and \textsc{Warmr}~\cite{Dehaspe1999}, 
\textsc{Foil}~\cite{Quinlan1990}, and \textsc{Progol}~\cite{Muggleton1995}, 
which are
frequent pattern learners. Both algorithms were implemented in the ACE Data Mining 
system \cite{ace}. The ACE system uses hipP \cite{hipp} as its
execution engine, a high-performance Prolog engine (written in C) with 
specific extensions for ILP such as the above mentioned query optimization
techniques.
A typical ILP benchmarks is the Mutagenesis data set~\cite{Srinivasan1996},
containing information about the structure of 230 molecules, and where the task of the
ILP system is to learn to predict whether an unseen molecule can cause
cancer or not.
A more real-life data set is the HIV data set \cite{DTP}, containing over 4000
examples.


\section{Gathering run-time information} \label{sec:traces}

Consider the generic ILP algorithm from Figure~\ref{fig:ilp}.
The target of query execution optimizations 
is the \textit{Evaluate} step, which takes a set of hypotheses to
be evaluated, and evaluates them on the current dataset. The other steps
that characterize the algorithm (such as finding suitable refinements for
queries) are not important from an engine implementor's point of view. However,
the latter are the most complex parts of the algorithm, and encompass most code
of the algorithm itself. 
For our debugging purposes, we extract enough information from an ILP run
necessary to be able to reproduce the \textit{Evaluate} step, without
running the ILP algorithm itself. More specifically, we only need to know the
queries that the algorithm runs, and on which example each query is evaluated. 
How and why these queries were generated and selected is irrelevant for
reconstructing the execution step.

\begin{figure}[tbp]
\begin{tabbing}
\p{query}((\p{atom}(\p{X},\p{'c'})), [1,2,3,4,5]). \\
\p{query}((\p{atom}(\p{X},\p{'h'})), [1,2,3,4,5]). \\
\\
\p{query}((\p{atom}(\p{X},\p{'c'}),\p{atom}(\p{Y},\p{'o'}),
    \p{bond}(\p{X},\p{Y})), [1,5]). \\
\p{query}((\p{atom}(\p{X},\p{'c'}),\p{atom}(\p{Y},\p{'c'}),
    \p{bond}(\p{X},\p{Y})), [1,5]). \\
\end{tabbing}
\caption{Example trace of an ILP run.}
\label{fig:example}
\end{figure}
To extract the desired information, we modify the \textit{Evaluate}
step from the ILP algorithm to record all evaluated queries to a file, which we
call the \emph{trace} of the algorithm. An example of such a trace 
after running a modified algorithm can be seen in 
Figure~\ref{fig:example}. This trace represents a run of an ILP 
algorithm that executed 4 queries: 2 queries that were executed on
all 5 examples, and 2 extensions of the first query, which were only executed 
on the first and the last example. 
Notice that this trace is no longer dependent of the concrete algorithm itself, in the
sense that it is just a
sequence of queries the algorithm evaluated on the examples.

\begin{figure}[tbp]
\begin{alltt}
\textit{% Run all queries from 'Trace' on 'Dataset'}
simulate(Trace, Dataset) :-
    read(Trace, Input),
    ( Input == end_of_file ->
        true
    ;
        Input = query(Query, Examples),
        evaluate_query(Examples, Query, Dataset),
        simulate(Trace, Dataset)
    ).

\textit{% Evaluate a query on a set of examples}
evaluate_query([], _, _).
evaluate_query([E|Es], Query, Dataset) :-
    load_example(Dataset, E),
    (call(Query), fail ; true),
    evaluate_query(Es, Query, Dataset).
\end{alltt}
\caption{\p{simulate/2}: A simple trace simulator.}
\label{fig:traces:simulator}
\end{figure}

The gathered information can now be run through a trace simulator which,
using the example database and background knowledge of the ILP algorithm,
can now simulate the execution step of the ILP algorithm. Such a trace
simulator is shown in Figure~\ref{fig:traces:simulator}, and does
nothing but run the original queries on the corresponding examples. While such
a simulator in itself can be used for manually debugging query execution,
we will also extend it further in Section~\ref{sec:ddebug} to yield an 
automatic debugging approach of different execution mechanisms.



\section{(Delta) debugging using traces} \label{sec:ddebug}

When developing optimizations for query evaluation, different execution 
mechanisms are investigated. When a new execution mechanism
should yield the same final results as the existing ones, inconsistencies can
be detected by running the ILP algorithm using each 
execution mechanism, and comparing the final results. 
For
example, for \textsc{Tilde}, one can compare the learned decision trees
to determine whether or not two runs are consistent with each other. 
This approach 
relies on the fact that new execution mechanisms speed up execution without
changing the final results of the ILP algorithm. 
However,
an inconsistent result only indicates that there is a bug in the 
execution \emph{somewhere}, but it doesn't show \emph{where}. To be able to 
determine this, the complete ILP algorithm has to be run using both 
the debugger of the host language of the ILP system (e.g. Prolog),
and the debugger of the host language of the execution
engine (e.g. C), where the actual bug of the execution mechanism is located. 
Because of the size and complexity of the ILP system, debugging
on both levels simultaneously is very hard and time-consuming in practice.
Moreover, testing execution mechanisms by comparing the output of the 
algorithm only works when the algorithm has 
deterministic behavior: if the decisions it makes are based on a random
factor, the outputs of the algorithm can (slightly) differ, and comparing
runs is not possible. This makes locating bugs in the implementation of
optimizations even harder.
Using execution traces for debugging solves many of these problems: trace 
execution is
deterministic, and a trace simulator is so small that the focus of the
debugging process is purely on the optimization itself. Moreover, traces
can speedup debugging even more drastically by limiting execution to
the part of the trace causing the bug, as we show in this section. \\

When two runs of a deterministic ILP algorithm produce different results, 
this means that the query evaluation process
selected different queries at some point. If the only difference between
both runs is a query optimization scheme, this means that the
optimization caused a query to succeed or fail where it did not before,
meaning a bug (assuming that optimizations preserve success or failure
of queries). Testing optimizations can therefore be 
reduced to comparing
the success of query with and without the optimizations scheme. This can be 
achieved by simply running the trace through a simulator that records
query successes, and runs every set of queries with and without the
optimization enabled. Not only can such a simulator detect bugs this way,
it can also pinpoint exactly in which query the bug occurs.

Due to the size of the trace, it might still be that a big
part of the execution needs to be analyzed to find the bug. A bug occurring
in a query is often also dependent on previously executed queries, 
which means that the trace cannot just be reduced to a single
query to be able to reproduce and locate the bug.
However, because the trace contains all the information determining the 
execution, locating a bug through traces can be turned into a 
\emph{data slicing}~\cite{Chan1998} problem. The goal of data slicing is to
take input data (i.e. a trace) that causes a bug to manifest itself, and
reduce this data as much as possible to yield a smaller subset of data
still exposing the bug. The standard approach to data slicing is simply to 
use binary search: split your data in two, test both halves, and 
continue with the half that still reproduces the bug. However, binary search
might be too coarse-grained to find a bug, and as such fail to reduce
the trace sufficiently. For example, if a bug occurs in the last query
of the trace because of the execution of the first query, neither of both
halves would reproduce the bug. \emph{Delta Debugging}~\cite{Zeller2002} is an
automated data slicing technique that overcomes these issues. We
describe delta debugging in the remainder of this section. \\

We briefly summarize the formalizations from~\cite{Zeller2002}.
Given a set of data $\mathcal{D}$ which causes a bug to appear. We
denote this as $\mathit{test}(\mathcal{D}) = \mathrm{fail}$.
$\mathcal{D}_g \subseteq \mathcal{D}$ is a \emph{global minimal data slice} 
if 
\[ 
    \mathit{test}(\mathcal{D}_g) = \mathrm{fail} \wedge 
    \forall \mathcal{D}' \subseteq \mathcal{D} \cdot 
        (|\mathcal{D}'| < |\mathcal{D}_g| \Rightarrow 
        \mathit{test}(\mathcal{D}') \neq \mathrm{fail})
\]
In other words, $\mathcal{D}_g$ is the smallest possible subset of the
original slice still reproducing the bug. Computing a global minimal data 
slice is infeasible in practice, since
it requires testing of all $2^{|\mathcal{D}|}$ subsets of $\mathcal{D}$,
which has exponential complexity. A less strict condition is the one of
the \emph{local minimum data slice} $\mathcal{D}_l$, for which no smaller 
subset exists that exposes the bug:
\[
    \mathit{test}(\mathcal{D}_l) = \mathrm{fail} \wedge 
    \forall \mathcal{D}' \subset \mathcal{D}_l \cdot 
    \mathit{test}(\mathcal{D}') \neq \mathrm{fail}
\]
However, testing whether $\mathcal{D}_l$ is indeed a local minimum still
requires $2^{|\mathcal{D}_l|}$ tests. An approximation to the local minimal
slice is an \emph{n-minimal data slice} $\mathcal{D}_n$, which is a slice for 
which no $n$ elements can be removed without making the bug disappear:
\[
    \mathit{test}(\mathcal{D}_n) = \mathrm{fail} \wedge 
    \forall \mathcal{D}' \subset \mathcal{D}_n \cdot
    ( |\mathcal{D}_n| - |\mathcal{D}'| \leq n \Rightarrow  
    \mathit{test}(\mathcal{D}') \neq \mathrm{fail})
\]
\begin{figure}[tbp]
\begin{center}
\begin{tabbing}
xxx \= xxx \= \hspace{6.5cm} \= \kill
{\bf function} \textsc{DDebug}($\mathcal{D}$) : \\
\> \textbf{return} \textsc{DDebug}($\mathcal{D}$,2) \\
\\
{\bf function} \textsc{DDebug}($\mathcal{D}$,$n$) : \\
\> $\Delta_{i=1}^{n}$ := \textsc{Partition}($\mathcal{D}$,$n$) \\
\> \textbf{if} $\exists \Delta_i, \mathit{test}(\Delta_i) = \mathrm{fail} $ : \\
\> \> \textbf{return} $\textsc{DDebug}(\Delta_i,2)$
\> \emph{\small -- `Reduce to subset'} \\
\> \textbf{else if} $\exists \Delta_i, \mathit{test}(\mathcal{D} - \Delta_i) = \mathrm{fail} $ :  \\
\> \> \textbf{return} $\textsc{DDebug}(\mathcal{D} - \Delta_i,\mathit{max}(n-1,2))$
\> \emph{\small -- `Reduce to complement'} \\
\> \textbf{else if} $n < |\mathcal{D}|$ : \\
\> \> \textbf{return} $\textsc{DDebug}(\mathcal{D},\mathit{min}(|\mathcal{D}|,2n))$ 
\> \emph{\small -- `Increase granularity'} \\
\> \textbf{else} : \\
\> \> \textbf{return} $\mathcal{D}$
\> \emph{\small -- `Done'}
\end{tabbing}
\end{center}
\caption{\textsc{DDebug}: The Delta Debugging algorithm. Finds a 1-minimal
    subset of $\mathcal{D}$ that causes the bug to appear.}
\label{fig:ddebug}
\end{figure}
The delta debugging algorithm~\cite{Zeller2002}, depicted in 
Figure~\ref{fig:ddebug}, finds a
1-minimal data slice of $\mathcal{D}$, i.e. a slice for which no one 
element can be removed without making the bug disappear. Note that even
smaller slices might be constructed by removing more than one element.
The algorithm works by dividing the data set in $n$ (more or less) equal
subsets, and checking if one of them still exposes the bug. If so, 
the process continues with this subset. If no subset exposes the bug
but a complement of one of the subsets does, the process continues with the
complement and increases granularity (such that the subsets in the next
step are equally large). Otherwise, the granularity is increased if
possible, or the process stops. 

\begin{figure}[tbp]
\begin{center}
\begin{tabular}{cclccccc}
\hline\hline
\textbf{Step} & \textbf{Call} & & \multicolumn{4}{c}{\textbf{Queries}} & \textbf{Result} \\
     & & & 1 & 2 & 3 & 4 \\
\hline
1 & \textsc{DDebug}($\{1,2,3,4\}$,2)
  & $\Delta_1$ & $\bullet$ & $\bullet$ &           &           & $\surd$ \\
& & $\Delta_2$ &           &           & $\bullet$ & $\bullet$ & $\surd$ \\
& & \multicolumn{6}{c}{\emph{Increase granularity}} \\
2 & \textsc{DDebug}($\{1,2,3,4\}$,4)
  & $\Delta_1$ & $\bullet$ &           &           &           & $\surd$ \\
& & $\Delta_2$ &           & $\bullet$ &           &           & $\surd$ \\
& & $\Delta_3$ &           &           & $\bullet$ &           & $\surd$ \\
& & $\Delta_4$ &           &           &           & $\bullet$ & $\surd$ \\
& & $\Delta_1^{-1}$ &      & $\bullet$ & $\bullet$ & $\bullet$ & $\times$ \\
& & \multicolumn{6}{c}{\emph{Reduce to complement}} \\
3 & \textsc{DDebug}($\{2,3,4\}$,3) \\
& & $\Delta_1$ &           & $\bullet$ &           &           & $\surd$ \\
& & $\Delta_2$ &           &           & $\bullet$ &           & $\surd$ \\
& & $\Delta_3$ &           &           &           & $\bullet$ & $\surd$ \\
& & $\Delta_1^{-1}$ &      &           & $\bullet$ & $\bullet$ & $\surd$ \\
& & $\Delta_2^{-1}$ &      & $\bullet$ &           & $\bullet$ & $\times$ \\
& & \multicolumn{6}{c}{\emph{Reduce to complement}} \\
4 & \textsc{DDebug}($\{2,4\}$,2) \\
& & $\Delta_1$ &           & $\bullet$ &           &           & $\surd$ \\
& & $\Delta_2$ &           &           &           & $\bullet$ & $\surd$ \\
& & \multicolumn{6}{c}{\emph{Done: $\{2,4\}$ is 1-minimal}} \\
\hline\hline
\end{tabular}
\end{center}
\caption{Example run of the delta debugging algorithm on a trace with 4
    queries.}
\label{fig:ddebugex}
\end{figure}
In our case, the data $\mathcal{D}$ corresponds to a trace, and 
every $\Delta_i$ represents a set of queries. Testing a $\Delta_i$
consists of running the trace with queries $\Delta_i$
through a trace simulator, and checking the output of the simulator
for consistent results. 
For example, suppose that we have a query trace with 4 queries exposing a
bug. Applying the delta debugging algorithm on the set of queries in the trace results
in the steps from Figure~\ref{fig:ddebugex}. 
Note that some tests are repeated: a smart implementation
can memorize tests, and re-use their answers.
An important factor that determines the speed of the trace slicing is
the granularity of the slicing process. Depending on what one considers the
smallest part in which a trace can be divided, the delta debugger needs to
consider more or less slices. A delta debugger for an ILP query trace can be 
set to use different granularities: it can
either choose to find failing iterations in a trace, 
which gives fast results, but also less compact traces;
it can prune the trace on the level of the queries themselves, giving
a minimal trace; and, it can trim down the number of examples on
which every query is run, reducing the number of times a query needs
to be called to expose a bug. For example, consider the trace of
Figure~\ref{fig:example}. If the delta debugger is set to find failing
iterations, it only needs to perform two tests, one for every iteration.
If it is set to find failing queries, it needs to consider each of the
four queries separately, which introduces more checks than only finding
the failing iteration. Finally, in the finest setting where every run of
a query is trimmed down, the delta debugger needs to consider the combinations
of the 14 runs (i.e. the first 2 queries are each run on 5 different example,
whereas the last 2 queries are run on 2 different examples).

In the worst case, the \textsc{DDebug} algorithm needs to perform
$|\mathcal{D}|^2 + 3|\mathcal{D}|$ tests. However, this worst case
seldom occurs in practice. In the optimal case where there is only one 
element in the slice causing the bug to appear, the number of tests is
bound by $2 \cdot log_2(|\mathcal{D}|)$. \\


\section{Implementation} \label{sec:implementation}

\begin{figure}[tbp]
\begin{center}
\includegraphics[scale=0.65]{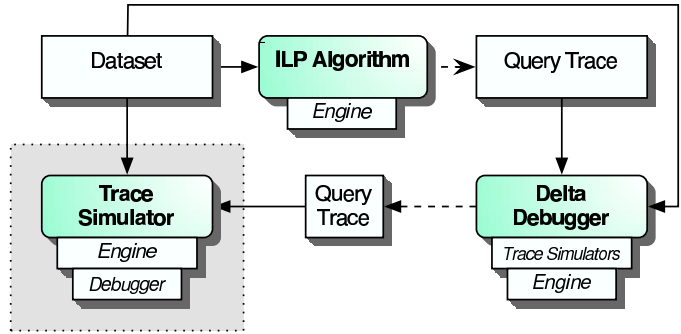}
\end{center}
\vspace*{-5mm}
\caption{Overview of the debugging process.}
\label{fig:traces:debugoverview}
\end{figure}
We have implemented and used the delta debugging approach 
in the development
of new execution mechanisms for the ACE Data Mining system. An overview
of the debugging process can be seen in Figure~\ref{fig:traces:debugoverview}.
The traces generated by the ILP algorithm are fed to the 
delta debugger, which trims it down to a smaller trace. 
The resulting trace is then fed into a trace simulator, and the engine
(i.e. hipP) can then be manually debugged using the host language debugger
(i.e. \texttt{gdb}).

We implemented two types of delta debuggers,
which differ in the type of test they perform to detect when
the execution of a trace exposes a bug. The simplest type of delta debugger
is one that runs a trace through a trace simulator run in a separate hipP
engine, and checks whether the process terminates successfully or not. This test
can be used for bugs that cause an engine to fail (e.g. due to a segmentation 
fault). The second type of test compares the trace execution of two engines
to check for inconsistent results. First, the queries from the original
trace are adorned with extra goals, recording for every query
in the trace on which
examples it succeeds. The test of the delta debugger then consists of
calling hipP and running the resulting trace through both a plain trace 
simulator (see Figure~\ref{fig:traces:simulator}) and 
a simulator with the (buggy) optimization enabled. The resulting logs
of both runs are compared, and if they differ, the test fails.
The delta debugger can be configured to use the different granularities
described in Section~\ref{sec:ddebug}: it can trim a trace to the
minimal number of failing iterations, to the minimal number of failing queries,
and, in its finest setting, to the minimal query runs (i.e. minimize both
the number of queries and the examples they are run on).

\begin{table}[tbp]
\begin{center}
\begin{minipage}{.8\textwidth}
\begin{tabular}{clrr|rrr}
\hline\hline
  \textbf{Trace}
& \textbf{Granularity}
& \textbf{Time}
& \textbf{Tests} 
& \multicolumn{3}{c}{\textbf{Resulting Trace}} \\
& & & 
    & It\footnote{Total number of iterations in the trace.} 
    & Qu\footnote{Total number of queries in the trace.}  
    & R\footnote{Total number of query runs necessary to reproduce the bug.} \\
\hline
1 & Iterations           &  16.2s &  10 & 1 & 137 &   822 \\
  & Queries           &  27.1s &  26 & 1 &   1 &     6 \\
  & Queries $\circ$ Iterations   &  18.9s &  24 & 1 &   1 &     6 \\
  & Examples $\circ$ Queries   &  27.6s &  29 & 1 &   1 &     1 \\
2 & Iterations           &  78.0s &  53 & 2 & 181 & 10942 \\
  & Queries           & 177.3s & 157 & 2 &   2 &   236 \\
  & Queries $\circ$ Iterations   & 120.0s & 136 & 2 &   2 &   236 \\
  & Examples $\circ$ Queries   & 180.4s & 171 & 2 &   2 &     2 \\
3 & Iterations           & 138.1s & 105 & 3 & 398 & 17235 \\
  & Queries           & 360.2s & 338 & 3 &   3 &   265\\
  & Queries $\circ$ Iterations   & 226.0s & 271 & 3 &   3 &   265 \\
  & Examples $\circ$ Queries   & 371.1s & 413 & 3 &   3 &     3 \\
\hline\hline
\end{tabular}
\end{minipage}
\end{center}
\caption{Delta debugger execution time and number of tests 
    performed
    for different granularities on three traces, together with
    statistics about the resulting traces. Traces are trimmed to the
    minimal amount of failing Iterations, Queries or
    Examples. Combinations of these granularities are denoted by
    $\circ$.}
\label{tab:traces:ddebug}
\end{table}
Table~\ref{tab:traces:ddebug} shows the execution time of the
delta debugger using different combinations of granularities.
For our experiments, we used a trace from a \Tilde\ run on the 
Mutagenesis data set, with a lookahead setting of 2. The trace consists
of 53 iterations of the algorithm, encompassing
a total of 12908 queries. This trace was modified to get three
variants: the first trace triggers an error when the last query of the last
iteration is executed; the second trace triggers the same bug, yet only if the first
query of the first iteration is executed as well; the third trace triggers 
the same bug whenever the first query 
and another query from the middle of the trace was executed. For
each of these traces, the delta debugger was run using different
granularities. Combinations of granularities are denoted
by $\circ$, where $G_1 \circ G_2$ means applying delta debugging with 
granularity $G_1$ on the trace resulting from delta debugging with 
granularity $G_2$.
The delta debugger successfully minimized all three traces to the minimal
trace needed to reproduce the bug, being a trace of 1, 2 and 3 queries
respectively.
The results show that applying the delta debugging first 
on the level of iterations, and then pruning further on the query level 
requires less tests than immediately pruning the complete 
trace on the query level. Pruning on the iteration level 
gives a first `rough' version of the trimmed down version of the trace, after 
which one can decide to prune further on the query level.


\section{Conclusion} \label{sec:conclusion}
In this work, we presented a trace-based approach to debugging
query execution mechanisms for ILP algorithms. Using traces
to perform debugging yields several advantages.
The specific workings of the ILP algorithm do not have to be known, as
the traces are algorithm independent, yet provide
enough information for performing a perfect simulation of the query execution 
of the algorithm itself. With trace-based execution, time is only spent on 
the execution of queries. Therefore, a complex query generation phase of an 
ILP algorithm does not affect the total execution time of a trace, and so 
debugging can be done faster. Finally, it is not necessary to have full
knowledge of the code base of the ILP system, which can in practice become
very large. 

By applying the delta debugging algorithm on traces, the 
number of queries can be reduced significantly, allowing bugs to be exposed 
very fast without having to manually step through the complete trace. \\

In the past, traces of execution have been used to understand 
misbehavior of programs \cite{Ducasse1999a,Ducasse1999b}. These approaches
do not use static traces, but instead interleave execution of the 
program with calls to the tracer, to avoid having to store the large
traces. In the context of debugging ILP query execution, not storing the
traces explicitly has the disadvantage that the execution times are
higher (because time is spent in the ILP algorithm itself), and the bug
might not occur if the algorithm is non-deterministic. Moreover, without
a static trace, applying delta debugging to reduce the total time needed
to expose a bug is not possible.


\bibliographystyle{abbrv}
\bibliography{debug}

\end{document}